\documentclass{PoS}

\title{A close look at compact triple radio sources }
\ShortTitle{A close look at compact triple radio sources }
\author{D\'avid Cseh$^a$, \'Ad\'am Fuhl$^a$, \speaker{S\'andor Frey}$^{b,c}$\thanks{This work has benefited from research funding from the European Community's sixth Framework Programme under RadioNet R113CT 2003 5058187, and from the Hungarian Scientific Research Fund (OTKA K72515). The EVN is a joint facility of European, Chinese, South African and other radio astronomy institutes funded by their national research councils. The US National Radio Astronomy Observatory is a facility of the NSF operated under cooperative agreement by Associated Universities, Inc.}\\
\llap{$^a$}E\"otv\"os University, Department of Astronomy, Budapest, Hungary\\
\llap{$^b$}F\"OMI Satellite Geodetic Observatory, Budapest, Hungary\\
\llap{$^c$}MTA Research Group for Physical Geodesy and Geodynamics, Budapest, Hungary\\
E-mail: \email{csehdavid@freemail.hu}, \email{afuhl@elte.hu}, \email{frey@sgo.fomi.hu}}

\abstract{We show an archival 1.4-GHz Very Large Array (VLA) image of the core-dominated triple (CDT) radio source J1628+3906, along with a recent high-resolution 5-GHz Very Long Baseline Interferometry (VLBI) image of its core. This core is resolved into two nearly symmetric components at $\sim10$ milli-arcsecond (mas) angular scale. The inner structure is markedly misaligned with respect to the large-scale structure. This morphology can be interpreted as a possible signature of once ceased and then restarted activity of the galactic nucleus, coupled with a rapid repositioning of the central radio jet axis. 
In a search for compact ($<5^{\prime\prime}$), optically identified, and highly redshifted ($z>3$) quasars in the Faint Images of the Radio Sky at Twenty-centimeters (FIRST) and the Sloan Digital Sky Survey (SDSS) catalogues, we have also found two quasars associated with weaker two-sided radio lobes at $\sim10^{\prime\prime}$ scale. We present archival 1.4-GHz VLA images of these quasars (J1036+1326 and J1353+5725). The compact cores of both of these CDT sources are likely to be detectable at mas scales. Future VLBI observations at the highest resolution could reveal if their inner radio structure is similar to that of the Compact Symmetric Objects (CSOs). The characteristic directions of their arcsecond-scale and mas-scale structures are probably misaligned which could also support the ``restarting and rapid repositioning'' scenario.} 

\FullConference{The 9th European VLBI Network Symposium on The role of VLBI in the Golden Age for Radio Astronomy and EVN Users Meeting\\
September 23-26, 2008\\
Bologna, Italy}

\begin{document}

A few extragalactic radio sources (active galactic nuclei, AGNs) exhibit triple structures on $\sim 100$~kpc projected linear scale with faint, nearly symmetric extended lobes, and relatively bright compact cores that appear unresolved ($<5^{\prime\prime}$) in the VLA FIRST survey \cite{Wh} at 1.4~GHz. This behaviour has been successfully interpreted as a possible signature of once ceased and then restarted activity of these {\it core-dominated triple} (CDT) sources, coupled with a rapid repositioning of the central radio jet axis \cite{Ma}. The change in the jet orientation results in a compact, beamed radio ``core'' which can even be detected with VLBI at three orders of magnitude smaller scales. On the other hand, the outer lobes are oriented nearly perpendicular to the line of sight. The periodic AGN activity may as well result in rare objects like double-double or triple-double radio galaxies \cite{Br}.
The concept of restarted activity may also work very well for X-shaped sources. This morphology of radio emission is believed to be produced by a sudden change in the outflow direction caused by a specific event (e.g. the capture of a secondary or merger black hole) in the evolution of a binary black hole system. They could be regarded as a parent population of CDTs. The existence of the latter class is simply the result of orientation and beaming effects on X-shaped sources \cite{Ma}. 


\begin{figure}[!h]
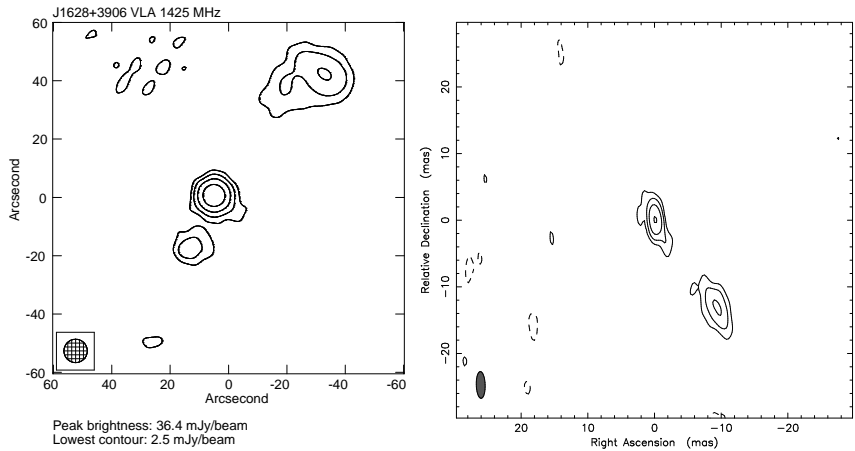

\centering
\includegraphics[bb=48 105 570 686, height=60mm, angle=0, clip=]{J1628+3906-VLA-BW.PS}
\includegraphics[bb=32 132 583 684, height=58mm, angle=0, clip=]{J1628+3906-VLBI.ps}
\caption{Radio images of J1628+3906. The positive contour levels increase by a factor of 2. The Gaussian restoring beams are shown in the bottom left corners. \textit{Left:} VLA, 1.4~GHz (project AM812, 2005 Apr 15). \textit{Right:} EVN, 5~GHz (project EF016, 2007 Mar 2). The peak brightness is 4.03~mJy/beam, the lowest contours are drawn at $\pm0.5$~mJy/beam. }
\label{fig1}
\end{figure}

We present an archival 1.4-GHz VLA B-array image of a CDT source (the galaxy J1628+3906), along with a high-resolution 5-GHz VLBI image of its core, which is unresolved with the VLA. The VLA image (Fig.~\ref{fig1}, left) shows an extended lobe structure on both sides of the core at arcminute scale. Based on a photometric redshift estimate ($z_{\rm photo} = 0.9$; SDSS DR6 \cite{Ri}), the projected linear size of the source is $\sim 600$~kpc. (A cosmological model with $H_{0}=70$~km~s$^{-1}$~Mpc$^{-1}$, $\Omega_{\rm m}=0.3$ and $\Omega_{\Lambda}=0.7$ is assumed.)
There are two weak ($\sim 2-4$~mJy) components detected at 5~GHz with the European VLBI Network (EVN) (Fig.~\ref{fig1}, right). The dominant direction of the inner structure -- two compact symmetric components or maybe a core--jet -- is markedly misaligned with respect to the large-scale structure. In a somewhat similar case of J1252+3310 \cite{Mo}, the source exhibits a complex, possibly curved inner jet structure. The comparison of the VLA and EVN images also reveals a structural misalignment. The observed structures in J1628+3906 and J1252+3310 can conveniently be reconciled with the ``restarting and repositioning'' scenario.

Motivated by the recent results of the Deep Extragalactic VLBI-Optical Survey (DEVOS; \cite{Mo,Fr}), we searched for compact high-redshift ($z>3$) radio sources that are unresolved ($<5^{\prime\prime}$) in the FIRST, with integral flux density $S_{\rm 1.4\;GHz}>20$~mJy, and optically identified with point-like quasars in the SDSS \cite{Sc}. The majority (nearly $90\%$) of such objects are potentially good candidates for VLBI detection at 5 GHz \cite{Fr}.\\

\begin{figure}[!h]
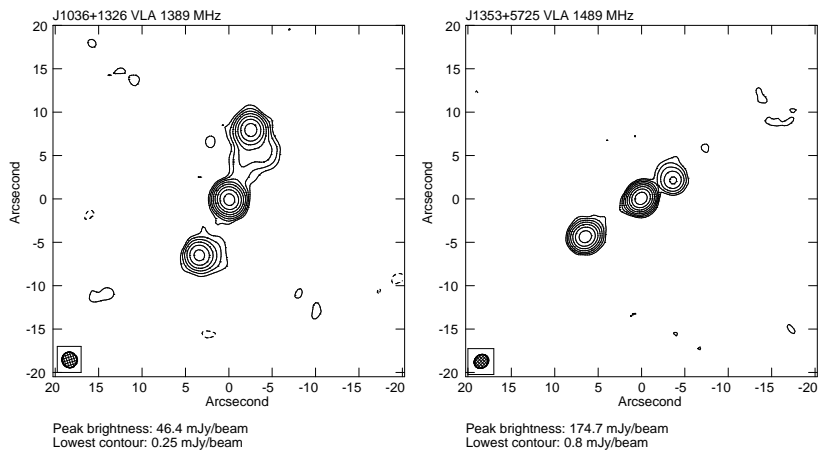

\center
\includegraphics[bb=48 105 570 686, height=60mm, angle=0, clip=]{1036+1326CONT.ps}
\includegraphics[bb=48 105 570 686, height=60mm, angle=0, clip=]{1353+5725CONT.ps}
\caption{1.4-GHz VLA images of two high-redshift quasars. \textit{Left:} J1036+1326, $z=3.10$, linear extent is 120~kpc (project AT126, 1991 Jul 8). \textit{Right:} J1353+5725, $z=3.46$, linear extent is 90~kpc (project AR250, 1991 Jun 28). The positive contour levels increase by a factor of 2. The Gaussian restoring beams are shown in the bottom left corners.}
\label{fig2}
\end{figure}

Surprisingly, we found two peculiar objects (J1036+1326 and J1353+5725) in our sample of $\sim 100$ compact high-redshift radio quasars. These two are associated with two-sided lobes (i.e. CDT sources). Although these objects are catalogued as unresolved in FIRST, our archival 1.4-GHz VLA A-array images (Fig.~\ref{fig2}) clearly reveal extended structures extending to about 100~kpc. We suspect that these are relics of previous activity in these quasars. Future VLBI observations could confirm if their mas-scale radio structure is compact, and its characteristic direction is misaligned with respect to the arcsecond-scale structure. In particular, their inner structure -- like in the case of J1628+3906 (Fig.~\ref{fig1}) -- could perhaps be similar to that of the Compact Symmetric Objects (CSOs) which are thought to be young radio sources. 


\end{document}